\documentclass[reprint,amsmath,amssymb,prl,lettersizepaper,tightenlines]{revtex4-1}

\usepackage{layout,float}
\usepackage{amsmath,amssymb}
\usepackage{graphicx}
\usepackage{dcolumn}
\usepackage{bm}
\graphicspath{{Images_article/}}

\newcommand{\be}{\begin{equation}}
\newcommand{\ee}{\end{equation}}

\newcommand{\bit}{\begin{itemize}}
\newcommand{\eit}{\end{itemize}}

\newcommand{\bra}[1]{\langle #1|}
\newcommand{\ket}[1]{|#1\rangle}

\begin{document}
\title{\large{Efficiently tunable photon emission from an optically driven artificial molecule}}

\author{Jonathan C. Lemus}
\affiliation{Grupo de F\'isica Te\'orica y Computacional, Escuela de F\'isica, Universidad Pedag\'ogica y Tecnol\'ogica de Colombia (UPTC), Tunja 150003, Boyac\'a, Colombia.}
\author{Hanz Y. Ram\'irez}
\email{hanz.ramirez@uptc.edu.co}
\affiliation{Grupo de F\'isica Te\'orica y Computacional, Escuela de F\'isica, Universidad Pedag\'ogica y Tecnol\'ogica de Colombia (UPTC), Tunja 150003, Boyac\'a, Colombia.}

\date{\today}

\begin{abstract}
In this work, we investigate the emission properties of double quantum dots driven by a monochromatic electromagnetic field, while undergoing resonant tunneling and dissipation by phonons. We found emission spectra for both cases, with and without phonons, whose comparison allows to determine the dissipative effects. On the basis of the obtained results we propose efficient control of the resonance fluorescence of an artificial molecule, suitable for optoelectronic applications.
\end{abstract}
\maketitle

\section{I. Introduction}

 The development of new technologies depends, to a large degree, on the use of structures with properties that optimize specific physical processes. In the area of  information, optical properties of quantum dots (QDs) are specially promising because their capability to confine electrons and electron-hole pairs (excitons). Important optical atomic properties, such as Rabi oscillations [1], photon antibunching [2], and resonance fluorescence (RF) [3], have been observed on QDs. For these reasons QDs are so-called artificial atoms. These semiconductor heterostructures are approximately 100 times bigger than natural atoms, and their properties can be reasonably controlled by means of growing techniques and gate technologies [4, 5]. 

QDs can be coupled by means of resonant quantum tunneling, constituting quantum dot molecules (QDMs), the so-called artificial molecules. QDMs are relatively new, and their properties are still a central point in many research groups around the world, on topics like their optical response. QDs, and QD-based systems differ from natural systems by characteristics proper of their solid state nature, such as phonon interaction presence. 

Experimental studies of resonance fluorescence of QDs and QDMs, at very low temperatures, can be carried out with ``negligible'' noise, leaving the radiative decay as the most important decay parameter. However, interaction with phonon reservoir or collisional processes have to be taken into account.

Even when important studies on QDM properties has been reported, it is necessary a deeper study on their RF. The theoretical research of Gao-xiang Li \textit{et al.} on the RF of a $\Lambda$-type three-level QDM in the Coulomb blockade regime, connected to two normal metal leads in a source-drain setting must be highlighted [6]. In their work, the intra-band optical transitions, and the effect of phonon dissipations on the RF were studied in the density matrix formalism, using the Fermi's golden rule to find the decay rate by the phonon channel.


In this article we present a theoretical optical characterization of an \textit{InGaAs/GaAs} QDM in the presence of an intense driving monochromatic electromagnetic field, interacting in the strong coupling limit, by means of its RF spectrum, and the effect of the phonon assisted emission on itself, to see how ``negligible'' is indeed this interaction for low temperatures. First, the interaction effect was studied, then we analyze its energy, and deduce the resulting eigenvalue equation that generates the so-called dressed states. Based on them, the allowed transitions are obtained to finally plot RF spectra for the artificial molecule under different conditions, including  phonon dissipation effects for various temperatures.  Phonon dissipation is thus observed in terms of the temperature dependence. We see how the case of a single QD can be studied from this particular QDM array. Finally the control of the properties of RF through the tunning of external parameters is proposed.

 \section{II. Coupled quantum dots}
 
Experimentally, QDs can be grown sufficiently near to give place to the overlapping of the wave functions of particles in the valence and conduction bands, so that these QDs interact, in pairs, by means of tunneling effect. Coupled quantum dots can be driven by a laser with frequency near to that of an exciton transition, as the frequency associated to the destruction of the direct exciton and corresponding emission of a photon.
 
 The system under study is composed of  two  quantum dots, coupled by quantum tunneling near to resonance (see figure 1). We will work in the Coulomb blockade regime, where the thermal energy of electrons is small compared to the energy associated with the addition of an electron in the QDM, i.e. $kT<E_{C}$, so that direct and indirect excitons are only considered in the setting, as is represented in figure 2. Direct exciton is placed in the left QD and the indirect one is disposed between the state in the valence band in the left QD and the state in conduction band of the right dot. As can be seen, energy difference between indirect exciton and direct exciton states ($E_{XD}$, and $E_{XI}$, correspondingly) is defined by $\Delta$. This QDM is driven by a monochromatic electromagnetic field, an intense field with frequency $\omega_{L}$ near to the frequency $\omega_{XD}$, associated to the direct exciton, i.e. it is a process in the strong interaction regime. 

The QDM-radiation states, with the radiation states in the Fock basis, are given as in figure 3. Besides this states, where we have no exciton state and $n+1$ photons ($\ket{n+1, g}$) in the basis too.

\begin{figure}
\includegraphics[scale=0.25]{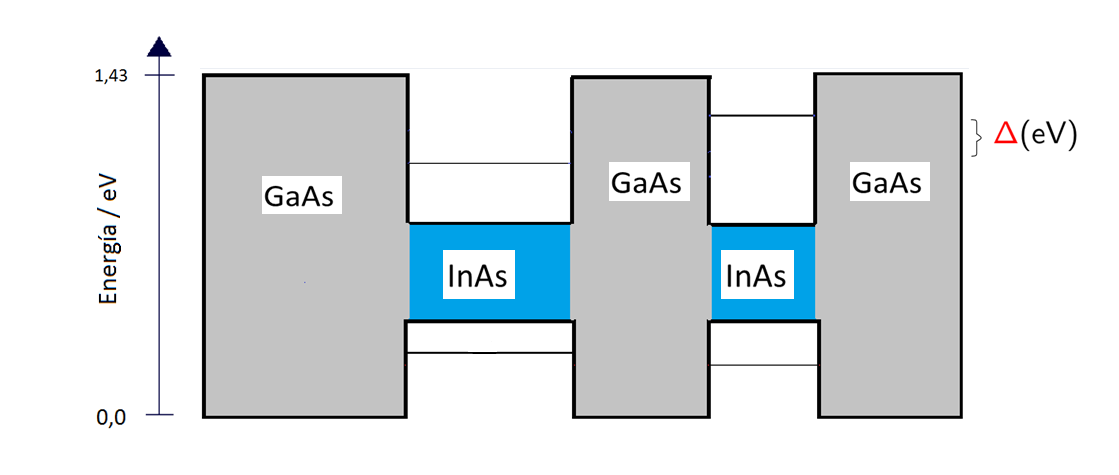}
\caption{General structure of the QDM at consideration.}
\end{figure}
 
Our research in on three-level QDMs. This distribution of states (figure 1) can be obtained with the appropriate quantization lengths for each QD and values of parameters into the composed system. Experimentally, a set of states, near in energy, is difficult to obtain, however. As we will see in subsection B, the detuning of exciton states can be tunned through an applied electric field, parallel to the indirect exciton state.

\begin{figure}
\includegraphics[scale=0.47]{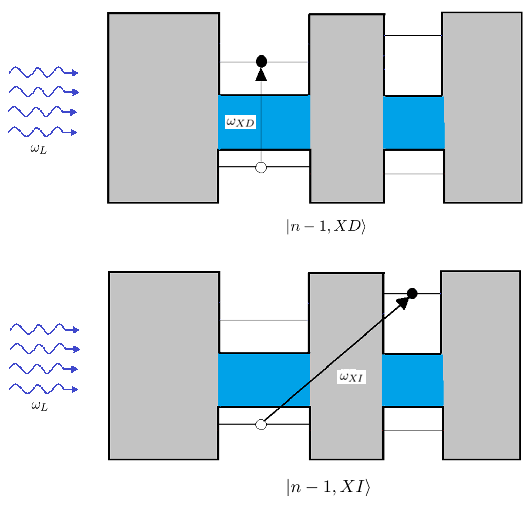}
\caption{Up: direct exciton state; down: indirect exciton state.}
\end{figure}

\subsection{A. Quantum processes in consideration.}

Due to the approximations done in our system, only some quantum processes were taken into account. Absorption of radiation from the light source, near to resonance with direct exciton, can happen for the creation of the direct exciton. The tunneling process keeps total angular momentum of the exciton, and consequently it occurs without emission of radiation, because conservation of angular momentum (photon possesses spin angular momentum). We assume exciton spin does not change, because we are interested in emission properties of the QDM. The decay of indirect exciton by radiative emission is negligible because its very small oscillator strength. 

Subsequently, when the phonon assisted emission of radiation is introduced, it assists photon absorption and emission, that creates and destroys direct exciton states as a mechanism of energy adjust when out of resonance processes are considered. In fact, phonon assisted tunneling process can be considered.

When the indirect exciton is generated (from the ground state by absorption of radiation and the tunneling process), after some time elapses, we can measure the emission of a phonon due to the tunneling process from the indirect state to the direct one, and a photon, produce in the decay of the of the direct state to the ground state.

\subsection{B. Manipulation of exciton states by an applied electric field.}

Direct and indirect excitons have characteristic energies $E_{XD}$ and $E_{XI}$, such that 

\begin{eqnarray}
\nonumber \hat{H}_{0}{\ket{XD}}&=&E_{XD}{\ket{XD}}, \\        \\
\nonumber\hat{H}_{0}{\ket{XI}}&=&E_{XI}{\ket{XI}},
\end{eqnarray}
where $\hat{H}_{0}$ is their  radiation-free Hamiltonian. Schematically, this situation is represented in figure 3, where $\Delta$ illustrates the difference between indirect exciton and direct exciton energies (Coulomb potential ignored for easiness of visualization)

\begin{equation}
\Delta=E_{XI}-E_{XD}.
\end{equation}
When a constant electric field, parallel to the growth direction is applied, it is possible to tune the indirect exciton energy without significantly changing  the direct exciton energy. Because of the applied field, the energy of the electron in the right dot is modified by

\begin{equation}
H=-q {d} \ {F}.
\end{equation}

\begin{figure}
\includegraphics[scale=0.45]{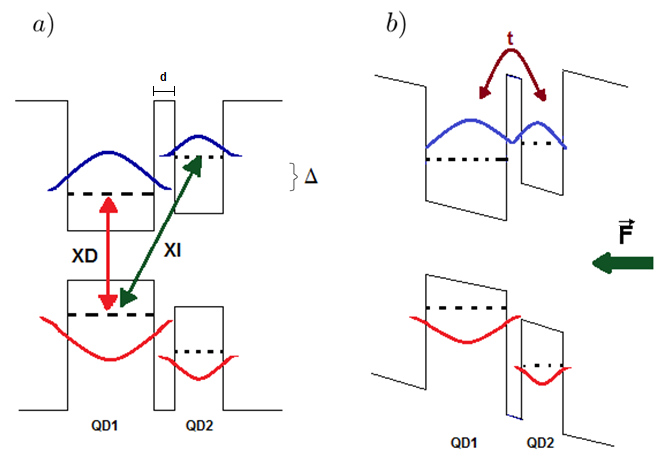} \\
\caption{Tunning scheme of the energy difference ($\Delta$) between indirect (XI) and direct (XD) energy states by means of an electric field $\vec{\textbf{F}}$. a) without the action of a field; b) with a field acting on the QDM.}
\end{figure}
 As a consequence, one can adjust the total energy of the exciton to a desired value. In the eigenvalue equations, the modification associated to the bias field reads
\begin{eqnarray}
\hat{H}_{0}{\ket{XD'}}&= &\hat{H}_{0}{\ket{XD}}=E_{XD}{\ket{XD}}, \\ \nonumber \\
\hat{H}_{0}{\ket{XI'}}&=&{E'_{XI}}{\ket{XI}}=\left( {E_{XI}-edE_{el}} \right) {\ket{XI}},
\end{eqnarray}
where $e$ is the electric charge, \textit{d} is the interdot distance (see figure 3), and $F$ is the electric field. As a consequence $\Delta$ can take any real value (within a physically reasonable interval). The new energy difference ${\Delta}_{n}$ is

\begin{equation}
\Delta_{n}=E'_{XI}-E'_{XD}=(E_{XI}-edE_{el})-E_{XD}.
\end{equation}
In particular, the bias field can be tuned, so that this difference is equal to zero

\begin{equation}
E_{XI}-edE_{el}=E_{XD}
\end{equation}

In this sense, the tunneling process can be enhanced, by means of the increasing of the correlation between direct and indirect states. Their wave functions overlap as energy distance between states diminishes (see figure 3). 

\subsection{C. Total Hamiltonian and energy level structure of the bare states}

The base of states we can work with is that of the tensor product of the QDM Hilbert space and that of the Fock space for the radiation. The energy level structure is shown in figure 4, where we have concentrated on the triplet of states

\begin{equation}
 \mathcal{E}(n-1) = \{ \ket{n, g}, \ket{n-1, XD}, \ket{n-1, XI}\}, 
 \end{equation}
 the so-called bare states of the QDM. 

The Hamiltonian for each identifiable triplet, including tunneling and radiation-matter interaction is

\begin{figure}
\includegraphics[scale=0.27]{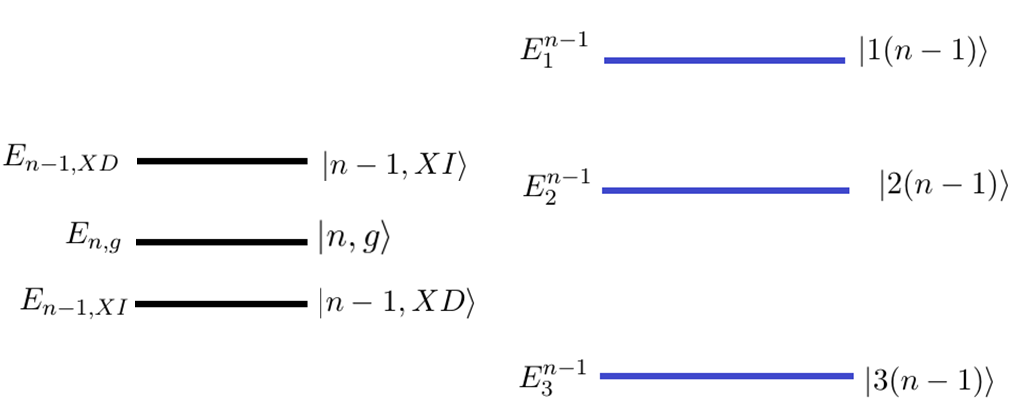}
\caption{Energy level structure. Left: bare states; right: dressed states.}
\end{figure}

\begin{equation}
\hat{H}_{T}=\hat{H}_{M}+\hat{H}_{R}+\hat{H}_{M-R}+\hat{H}_{t},
\end{equation}
where

\begin{eqnarray}
\nonumber \hat{H}_{M}&=&{E_{XD}}\ket{n-1,XD}\bra{n-1,XD}  \\  \\
\nonumber &+&{E_{XI}}\ket{n-1,XI}\bra{n-1,XI}+E_{0}\ket{n,g}\bra{n,g}, 
\end{eqnarray}
is the free-QDM Hamiltonian, 
\begin{eqnarray}
\nonumber \hat{H}_{R}&=&{n \hbar \omega_{L}} \ket{n,g}\bra{n,g}+{(n-1) \hbar \omega_{L}} \ket{n-1,XD}\bra{n-1,XD} \\
 &+& \ {(n-1) \hbar \omega_{L}} \ket{n-1,XI}\bra{n-1,XI},
\end{eqnarray}
the free-radiation Hamiltonian,
\begin{equation}
\hat{H}_{M-R} = g{\sqrt{n}} \ket{n,g}\bra{n-1,XD}+g{\sqrt{n}}\ket{n-1,XD}\bra{n,g},
\end{equation}
the radiation-QDM interaction Hamiltonian, and 
\begin{equation}
\hat{H}_{t}= t \ket{n-1,XD}\bra{n-1,XI}+t \ket{n-1,XI}\bra{n,XD},
\end{equation}
is the tunneling Hamiltonian between exciton states. Here we have defined $\Delta$ as the energy difference between direct and indirect exciton states, as before. $E_{0}$ is the energy of the radiation field, where there is no exciton state. $\omega_{L}$ is the frequency of the radiation source. $g$ is the radiation-matter coupling constant. $t$ is the tunneling rate. 

Bare states are eigenstates of the non-interacting radiation-QDM composed system, however, it is easy to see that these are not eigenvalues of $\hat{H}_{T}$, due to the tunneling and radiation-QDM interaction, which become off-diagonal terms in the corresponding matrix representation.

\subsection{D. Dressed states}

To solve the eigenvalue problem of $\hat{H}_{T}$, we focus on the triplet $\mathcal{E}(n-1)$ (the extended version of the doublet in RF of two-level QDs).

 After diagonalizing $\hat{H}_{T}$ Hamiltonian, we found eigenstates which are the superposition of bare states in $\mathcal{E}(n-1)$. Their eigenvalues are functions of $t$, $g\sqrt{n}$, and $\Delta$. These states are the corresponding dressed states of the artificial molecule, defined by the equations ( in compact notation) 
\begin{eqnarray}
\nonumber \ket{i(n-1)}&=&C_{g}^{i,n-1} \ket{n,g}+C_{XD}^{i,n-1}\ket{n-1,XD} \\ 
&+& C_{XI}^{i,n-1}\ket{n-1,XI},
\end{eqnarray}
for $i=1,2,3$, and $n-1$ the number of photons in the exciton states (see Appendix); energy eigenvalues were computed, together with the superposition coefficients, $C_{g}^{i,n-1}$, $C_{XD}^{i,n-1}$, and $C_{XI}^{i,n-1}$ [7]. The new energy structure,(eigenvalues of the dressed states) is represented schematically in figure 4, in comparison to energy of bare states. Energy of dressed states is denoted by $E_{1}^{n-1}$, $E_{2}^{n-1}$, and $E_{3}^{n-1}$. Fixing all the variables except the exciton energy difference $\Delta$, we plot these energies as can be seen in figure 5.

\begin{figure}
\includegraphics[scale=0.4]{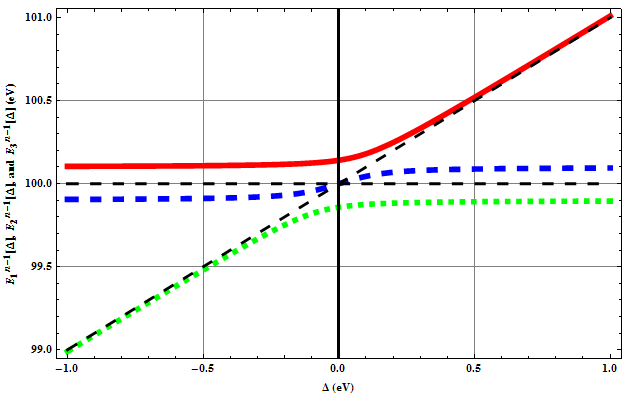}
\caption{Graphic representation of the energies of the dressed states on the exciton energy difference $\Delta$}
\end{figure}  

\subsection{E. Phonon dissipation}

To introduce dissipation in the RF spectrum we use a phenomenological model. Phonons are quantized vibrations of the lattice, and its energy determine the temperature of the solid. In the regime of zero temperature limit, for example, these vibrations are negligible. Now, for $T \neq 0$, the influence of phonons can be elucidated though the temperature dependence of RF. 

In experimental studies carried out by  M. Bayer and A. Forchel, and C. Kammerer \textit{et al.}, a phenomenological relation between the population decay rate of the exciton and temperature was found [8, 9]. This relation is given as

\begin{equation}
\Gamma (T)=\Gamma_{0}+a \ T+b \ e^{-\Delta E/kT},
\end{equation} 
where $\Gamma_{0}$ is the zero temperature line-width, and $a \ T$ and $b \ e^{-\Delta E/kT}$ are contributions from the acoustic-phonon and optical-phonon interactions, respectively. The acoustic-phonon term is linear in temperature and constitutes the greater contribution for $T \leq 40 K $. Optical-phonon contribution is exponential and is important at higher temperatures. For our study at low temperatures it is sufficient to consider the phonon effects in the linear approximation (acoustical phonon interaction). So we take

\begin{equation}
\Gamma (T)=\Gamma_{0}+a \ T.
\end{equation} 

According to the publications reporting this temperature dependence, the acoustic-phonon broadening efficiency $a$ must be of the same order than  the radiative decay rate $\Gamma_{0}$. All this in concordance with low temperature experimental researches [10].


\section{III. Resonance fluorescence}

\subsection{A. Allowed transitions}

Having diagonalized the total Hamiltonian $\hat{H}_{T}$, we find the eigenvalues and consequently the corresponding transition energies between dressed states in contiguous triplets (for the limit of an intense radiation source, i.e. $n \gg 1$). The allowed transitions depends on the value of $\Delta$ (leaving all the other parameters fixed). These are defined by

\begin{eqnarray}
\nonumber E_{ij}&=&E_{i}^{n-1}-E_{j}^{n-2} \\
&=&E_{i}^{n-1}-(E_{j}^{n-1}-\hbar \omega_{L}),
\end{eqnarray}
 where subindex $i$ gives the number of dressed state in the upper triplet (the initial state), and index $j$ gives the dressed state of the lower triplet (the final state).

This $\Delta$-dependence of the transition energies is shown in figure 6. When $\Delta$=0 eV, there are five possible transitions, when $\Delta$= 0.1 eV, there are seven possible transitions, and when $\Delta \gg 0$ there are just three reachable transition energies. The first two cases are represented schematically in figure 7 and figure 8. In the case of $\Delta \neq 0$ each color represents a different transition with a specific energy. We can see how the number of transition energies depends on the distance between energy levels in each triplet.

\begin{figure}
\includegraphics[scale=0.42]{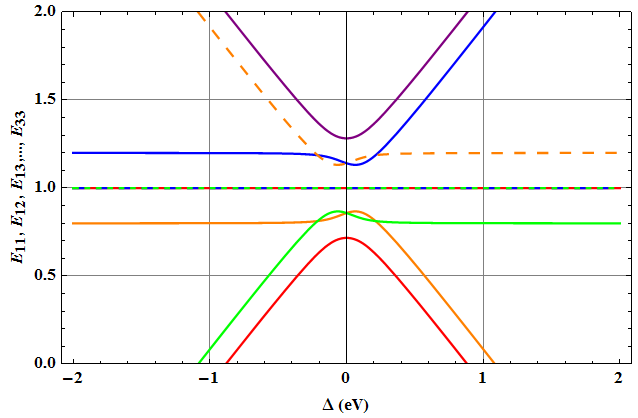}
\caption{Transitions energies $E_{ij}$ as functions of $\Delta$.}
\end{figure}

\begin{figure}
\includegraphics[scale=0.35]{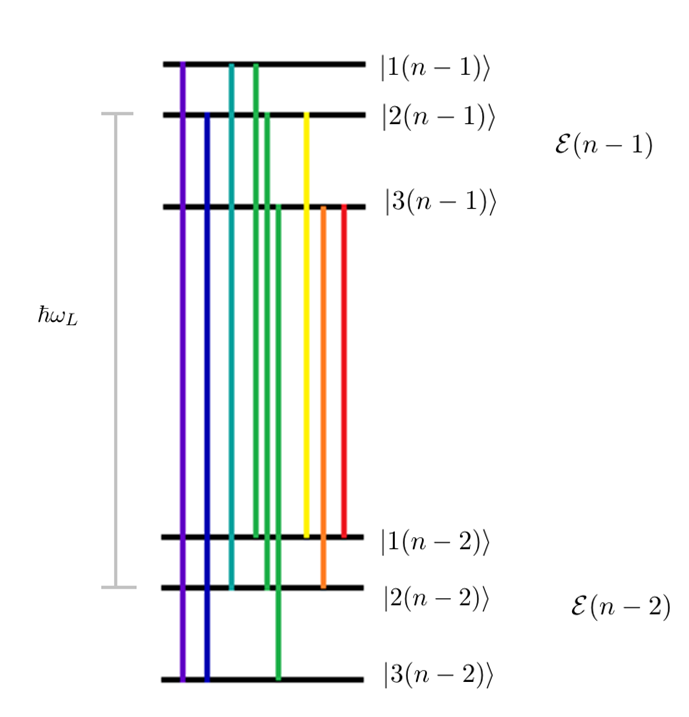}
\caption{Possible transitions between triplets $\mathcal{E}(n-1)$ and $\mathcal{E}(n-2)$, for $\Delta\neq 0$.}
\end{figure}

\begin{figure}
\includegraphics[scale=0.35]{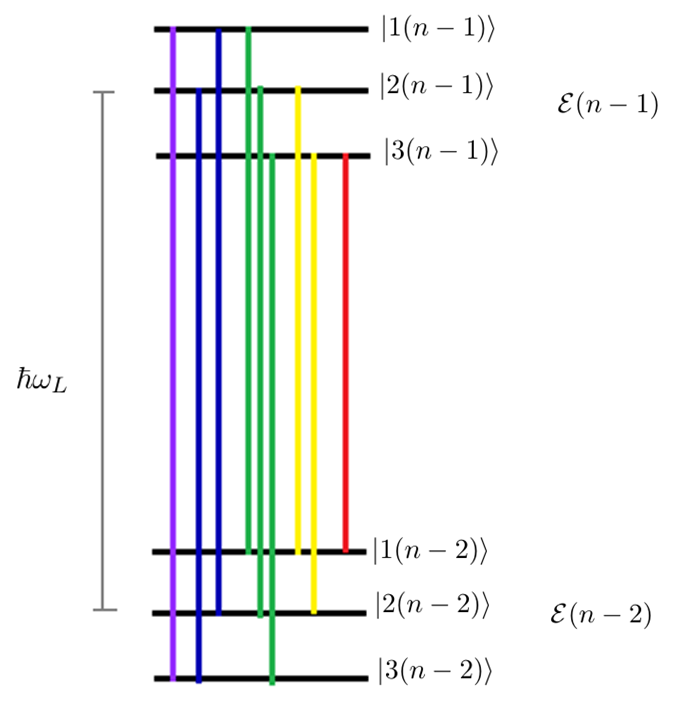}
\caption{Possible transitions between triplets $\mathcal{E}(n-1) and \mathcal{E}(n-2)$, for $\Delta= 0$.}
\end{figure}

\subsection{B. Algebraic form of the RF spectrum}

The resonance fluorescence of  an artificial molecule can be analyzed in detail beginning from the two-level quantum dot case. To find an algebraic expression for the RF spectrum of a driven QDM, we can appeal to the phenomenology of the driven single dot case.

It is well known that  emission for a quantized system, in despite of exhibiting completely discrete states, does not correspond to a perfect delta function, because fluctuations and finite lifetimes of quantum states modify the specific energy transitions, resulting in a Lorentzian function. Now an important question arise: how can we find the specific characteristic parameters of that Lorentzian distribution function for our case?

The unnormalized  Lorentz distribution and centered in b, is defined as

\begin{equation}
S(x)= A \frac{{w}^2}{(x -b)^2 + {w} ^2},
\end{equation} 
where A is the amplitude of the distribution, w is the half width at half maximum (HWHM), and b is the statistical mean value. Having this in mind we proceed to find the general expression for the RF spectrum. 

For the case of RF, where such distribution represents photon emission from optical transitions, A must be equal to intensity, b is the position of the peak and $\Gamma$ and $\gamma$, the population rate decay and radiative decay rate, respectively, must be related to $w$.

It is important to consider that central peak has a different behavior than that of lateral [11]. The general structure of a spectral peak is

 \begin{equation}
S(\Delta ')= I_{ij} \frac{\left[ {f(\Gamma, \gamma)} \right] ^2}{(\Delta ' -a_{ij})^2 + \left[ f({\Gamma, \gamma)} \right] ^2},
\end{equation} 
where $\Delta '$ is defined as $\Delta ' \equiv  \hbar (\omega - \omega_{L})$, with $\omega$ as the frequency of the spontaneously emitted radiation, $I_{ij}$ is the intensity of the corresponding optical transition, from dressed state $i$ in a triplet and $j$ is the final one in the lower triplet, and $a_{ij}$ is the corresponding transition energy. In turn $f(\Gamma, \gamma)$ is the HWHM function, where $\Gamma$ represents all the dissipation channels of the system and $\gamma$ is a pure radiative decay constant. Integrated luminosity depends on the transition amplitude, which is related to the squared particular dipole moment matrix element. For the intensities we have

\begin{equation}
L_{ij} \varpropto |\bra{j(n-2)} {\hat{\textbf{d}}} \ket{i(n-1)}|^2 \equiv |d_{i,j}|^2,
\end{equation}
for $ i,j=1,2,3$, for the intensity of each peak we have

\begin{equation}
I_{ij} \varpropto \frac{L_{ij}}{f_{ij}(\Gamma, \gamma)},
\end{equation}

Based on this proportionality we can find the relative intensity for every transition by calculating this matrix elements. These quantities in terms of the superposition coefficients given by

\begin{equation}
|d_{i,j}|=\mu |\ C_{g}^{i,n-2} C_{XD}^{j,n-1}|,
\end{equation}
where $\mu$ is the dipole momentum constant. All that we have written so far is valid for all the peaks. Finally, we need to find the particular form of the HWHM function. A reasonable first approximation to this function for the side peaks is the average value of $\Gamma$ and $\gamma$, i.e.,

\begin{equation}
f_{sp}(\Gamma, \gamma)=\frac{\Gamma+\gamma}{2}.
\end{equation} 
For the central peak we follow the approximation taken in references [12, 13], which yields

\begin{equation}
f_{cp}(\Gamma, \gamma)=\frac{\Gamma}{2}.
\end{equation} 

With all these we are able to plot our RF spectrum. Summarizing, for the side peaks, the corresponding spectrum is
 
\begin{equation}
S(\Delta ')= I_{ij} \frac{\left( {\frac{\Gamma+\gamma}{2}} \right) ^2}{(\Delta ' -a_{ij})^2 + \left( {\frac{\Gamma+\gamma}{2}} \right) ^2},
\end{equation} 
for $i \neq j$. And for the central peak,

\begin{equation}
S(\Delta ')= I_{ii} \frac{\left( {\frac{\Gamma}{2}} \right) ^2}{{\left(\Delta ' -a_{ii} \right)}^2 + \left( {\frac{\Gamma}{2}} \right) ^2}.
\end{equation}




\section{IV. Results}

\subsection{A. Resonance fluorescence spectra}

With the algebraic form of the RF of a QDM, and a $\Delta$ given, we can find the corresponding parameters, as peaks positions according to figure 6, and superposition constants for the dipole strength, to finally plot the spectrum. The first step is to check the validity of the model. This is done by means of the expectable result of the Mollow triplet when $\Delta$ is sufficiently large to make the coupling between the QDs negligible. For this case we obtain the spectrum shown in figure 9. The values we used are as follows

\begin{eqnarray}
\nonumber n&=&100 \ (number \ of \ photons), \\
\nonumber g\sqrt{n}&=&0.1 \ eV, \\
 t&=&0.1 \ eV, \\
\nonumber E_{XD}&=&1 \ eV, \\
\nonumber \hbar \omega_{L}&=&1 \ eV.
\end{eqnarray}

In the same way, for the case of $\Delta=0$ \ eV, the resultant spectrum is the structure of five peaks shown in figure 10. Finally, for the case of $\Delta = 0.008$ \ eV, the spectrum is a seven peaks structure, as shown in figure 11.
\begin{figure}
\includegraphics[scale=0.55]{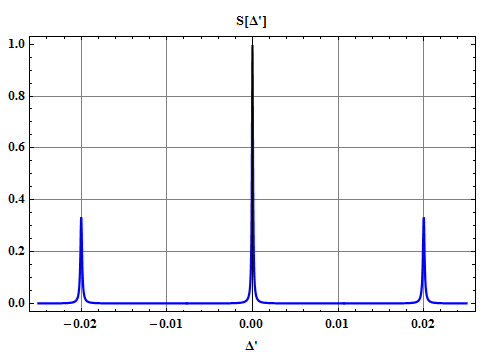}
\caption{Three-peaks spectrum}
\end{figure}

\begin{figure}
\includegraphics[scale=0.55]{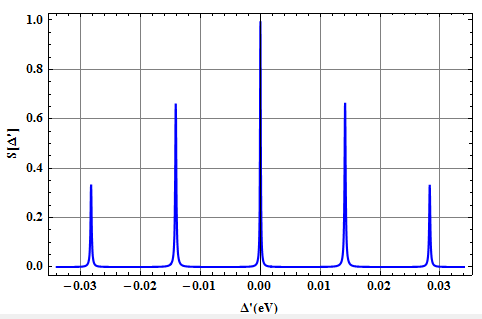}
\caption{Five-peaks spectrum}
\end{figure}

\begin{figure}
\includegraphics[scale=0.38]{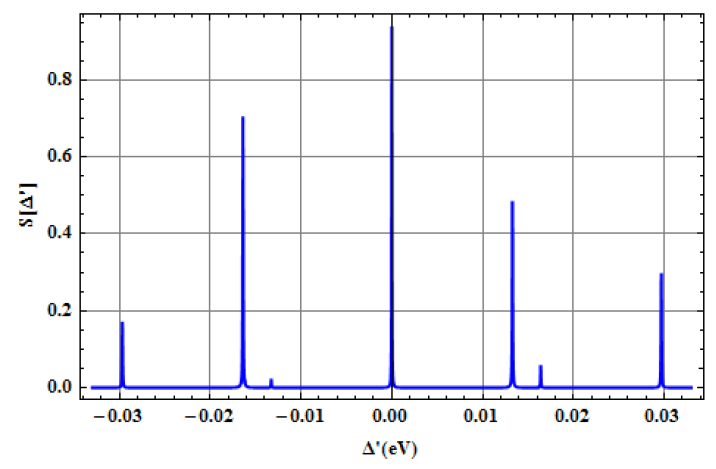}
\caption{Seven-peaks spectrum}
\end{figure}

\subsection{B. Phonon dissipation effects on the resonance fluorescence spectra}

Introducing the temperature dependent population decay rate (equation (16)), into the expression for the spectrum, together with the values $\Gamma_{0}=75 \mu$eV, and $a=22 \mu$eV/K, temperature dependent RF spectra can be obtained. When  $\Delta=0.008$ eV, we plot the spectrum at three different temperatures: $T=5 $ K, where the effects are small, but we can appreciate a small spectrum broadening in all the peaks. The resulting effect is described by spectrum in figure 12, where a diminution in the height of the peaks is notorious, and a small broadening of the peaks is appreciable.

\begin{figure}
\includegraphics[scale=0.65]{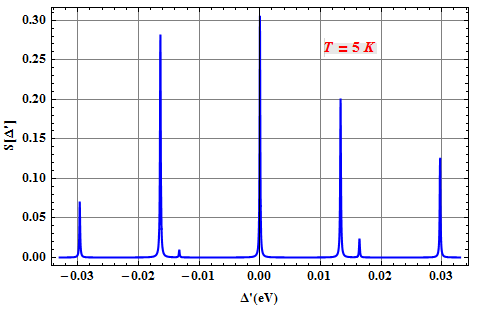}
\caption{RF spectrum of a QDM for $\Delta= 0.008$ \ eV and $T=5$ K.}
\end{figure}

For T=20 K, the diminution in the hight of the peaks is clear and the broadening grater, as can be appreciated in figure 13. This is a logical result, after all the inverse of the broadening of the spectrum is equivalent to the lifetime of the exciton. As the phonon decay channel increases, the lifetime of the exciton decreases. The last case, where T=40 K, is shown in figure 14.

\begin{figure}
\includegraphics[scale=0.65]{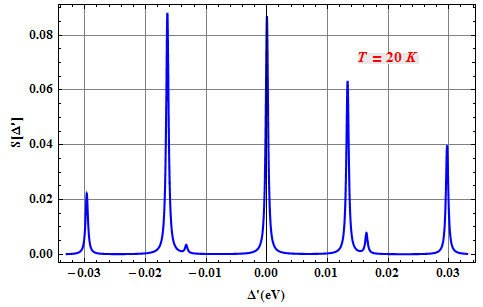}
\caption{RF spectrum of a QDM for $\Delta= 0.008$ \ eV and $T=20$ K.}
\end{figure}

\begin{figure}
\includegraphics[scale=0.65]{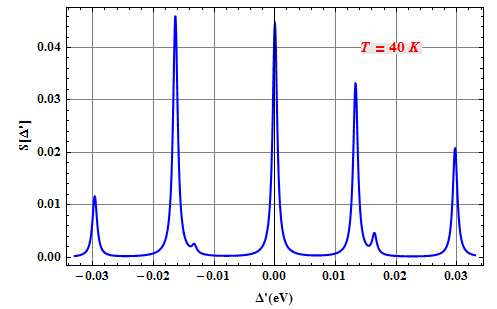}
\caption{RF spectrum of a QDM for $\Delta= 0.008$ \ eV and $T=40$ K.}
\end{figure}

\begin{figure}
\includegraphics[scale=0.55]{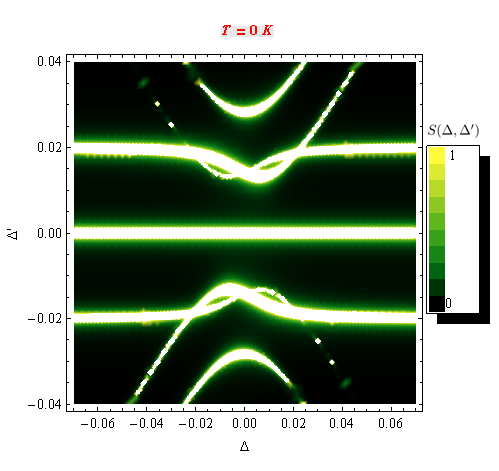}
\caption{Spectrum of the RF as a function of the QDM-radiation detuning energy ($\Delta'$) and exciton energy difference($\Delta$).}
\end{figure}

As the more general case, the spectrum of intensity as a function of $\Delta'$ and the exciton energy difference for T=0 K is given in figure 15. The different cases we have found previously can be observed, with the spectrum of five peak in the center, the spectrum of seven peaks for values near to 0.01 eV, and the Mollow triplet for values grater than 0.05 eV.




\section{V. Summary and conclusions}

Resonance fluorescence of an artificial molecule under the influence of a monochromatic laser has been studied. Emission spectra in the strong radiation-matter coupling regime were obtained. Effects of dissipation by phonons were considered by incorporating temperature dependence into such spectra and their influence then evidenced. As main conclusions from this study we could mention:

\begin{itemize}

\item{} The Jaynes-Cumming rungs in the studied system are composed by triplets instead of doublets as in the case of a driven single dot. 

\item[$\bullet$] It is possible to tune the exciton energy difference by means of an external electric field, in such a way that depending on the resulting $\Delta$, one can obtain a seven-peaks, five-peaks, or three-peaks spectra .

\item[$\bullet$] For a big exciton energy difference it is possible to recover the Mollow triplet.

\item[$\bullet$] Phonon interaction in artificial molecules generates a significant reduction in the exciton lifetime, represented by the broadening of peaks in the resonance fluorescence spectrum.

\item{} Even when tunneling rate is a fixed parameter of the system, the radiation-matter coupling can be varied by means of the number of photons so that additional control on the optical response of the system can be gained.

\item{} A future study, using Winier-Khintchine theorem, and density matrix theory to find, within a polaron framework, the power spectrum in presence of phonons would be in order as a next step in the study of this problem.

\end{itemize}


\section{Appendix}

\begin{eqnarray}
\nonumber \ket{1(n-1)}&=&C_{g}^{1,n-1} \ket{n,g}+C_{XD}^{1,n-1}\ket{n-1,XD} \\
\nonumber &+& C_{XI}^{1,n-1}\ket{n-1,XI}, \\  \nonumber \\
\nonumber \ket{2(n-1)}&=&C_{g}^{2,n-1} \ket{n,g}+C_{XD}^{2,n-1}\ket{n-1,XD} \\
\nonumber &+& C_{XI}^{2,n-1}\ket{n-1,XI}, \\ \\
\nonumber \ket{3(n-1)}&=&C_{g}^{3,n-1} \ket{n,g}+C_{XD}^{3,n-1}\ket{n-1,XD} \\
\nonumber &+& C_{XI}^{3,n-1}\ket{n-1,XI},
\end{eqnarray}





\section{Acknowledgements}
The authors acknowledge the Research Division of Universidad Pedag\'ogica y Tecnol\'ogica de Colombia for financial support.

\bibliography{References}

\begin{thebibliography}{190}

\bibitem{Htoon} H. Htoon, T. Takagahara, D. Kulik, O. Baklenov, A. L. Holmes, Jr., and C. K. Shih, \textit{Phys. Rev. Lett.} \textbf{22}, 087401 (2002).

\bibitem{Press}  D. Press, S. Gotzinger, S. Reitzenstein, C. Hofmann, A. Loffler, M. Kamp, A. Forchel, and Y. Yamamoto, \textit{Phy. Rev. Lett.} \textbf{98}, 117402 (2008).

\bibitem{Muller}  A. Muller, E. B. Flagg, P. Bianucci, X. Y. Wang, D.G. Deppe, W. Ma, J. Zhang, G. J. Salamo, M. Xiao, and C. K. Shih, \textit{Phy. Rev. Lett.} \textbf{99}, 187402 (2008).

\bibitem{Bimberg}   D. Bimberg, M. Grundmann, and N. N. Ledenstov, (\textit{Quantum Dot Heterostructures}, Jhon Wiley  Sons, 1999, England, Chapter 2).

\bibitem{Kemerink}  M. Kemerink, and L. W. Molenkamp, \textit{Appl. Phys. Lett.} \textbf{65}, 1012 (1994).

\bibitem{Li} G.-x. Li, S.-p. Wu, and J.-p. Zhu, \textit{J. Opt. Soc. Am. B} \textbf{27}, 1634 (2010).

\bibitem{Numerical} Numerical calculations were carried out by using \textit{Mathematica 8.0}.

\bibitem{Bayer} M. Bayer, and A. Forchel, \textit{Phys. Rev. B} \textbf{65}, 041308(R) (2002).

\bibitem{Kammerer} C. Kammerer, C. Voisin, G. Cassabois, C. Delalande, P. Roussignol, F. Klopf, J. P. Reithmaier, A. Forchel, and J. M. G$\grave{e}$rard, \textit{Phys. Rev. B} \textbf{66}, 041306(R) (2002).

\bibitem{Fujisawa} T. Fujisawa, T. H. Oosterkamp, W. G. van der Wiel, B. W. Broer, R. Aguado, S. Tarucha, and L. P. Kouwenhoven, \textit{Science} \textbf{282}, 932 (1998).

\bibitem{Vamivakas} A. N. Vamivakas, Y. Zhao, C.-Y. Lu, and M. Astat\"ure, \textit{Nature} \textbf{5}, 198 (2009).

\bibitem{Wei} Y.-J. Wei, Y. He, Y.-M. He, C.-Y. Lu, J.-W. Pan, C. Schneider, M. Kamp, S. H\"ofling,
D. P. S. McCutcheon, and A. Nazir, \textit{Phys. Rev. Lett.} {\bf 113}, 097401 (2014).

\bibitem{Karwat} P. Karwat, A. Sitek, and P. Machnikowski, \textit{Phys. Rev. B} \textbf{84}, 195315
(2011).


\end{thebibliography}


\end{document}